\documentclass[twocolumn]{aastex631}
\usepackage{amssymb,amsmath}

\newcommand{\mjysr}{MJy sr$^{-1}$}

\received{Oct 21, 2022}
\revised{Dec 29, 2022}
\submitjournal{The Astrophysical Journal Letters}
\shorttitle{PAH 3.3 Map}
\shortauthors{Sandstrom et al.}

\graphicspath{{./}{figures/}}

\begin{document}

\title{PHANGS-JWST First Results: Mapping the 3.3 \micron\ Polycyclic Aromatic Hydrocarbon Vibrational Band in Nearby Galaxies with NIRCam Medium Bands}

\correspondingauthor{Karin Sandstrom}
\email{kmsandstrom@ucsd.edu}

\author[0000-0002-4378-8534]{Karin M. Sandstrom}
\affiliation{Center for Astrophysics \& Space Sciences, Department of Physics, University of California, San Diego, 9500 Gilman Drive, San Diego, CA 92093, USA}

\author[0000-0002-5235-5589]{J\'er\'emy Chastenet}
\affil{Sterrenkundig Observatorium, Ghent University, Krijgslaan 281-S9, 9000 Gent, Belgium}

\author[0000-0002-9183-8102]{Jessica Sutter}
\affiliation{Center for Astrophysics \& Space Sciences, University of California, San Diego, 9500 Gilman Drive, San Diego, CA 92093, USA}

\author[0000-0002-2545-1700]{Adam~K.~Leroy}
\affiliation{Department of Astronomy, The Ohio State University, 140 West 18th Avenue, Columbus, OH 43210, USA}

\author[0000-0002-4755-118X]{Oleg V. Egorov}
\affiliation{Astronomisches Rechen-Institut, Zentrum f\"{u}r Astronomie der Universit\"{a}t Heidelberg, M\"{o}nchhofstra\ss e 12-14, 69120 Heidelberg, Germany}

\author[0000-0002-0012-2142]{Thomas G. Williams}
\affiliation{Sub-department of Astrophysics, Department of Physics, University of Oxford, Keble Road, Oxford OX1 3RH, UK}
\affiliation{Max-Planck-Institut f\"ur Astronomie, K\"onigstuhl 17, D-69117 Heidelberg, Germany}

\author[0000-0002-5480-5686]{Alberto D. Bolatto}
\affiliation{Department of Astronomy and Joint Space-Science Institute, University of Maryland, College Park, MD 20742, USA}

\author[0000-0003-0946-6176]{Médéric~Boquien}
\affiliation{Centro de Astronomía (CITEVA), Universidad de Antofagasta, Avenida Angamos 601, Antofagasta, Chile}

\author[0000-0001-5301-1326]{Yixian Cao}
\affiliation{Max-Planck-Institut f\"{u}r Extraterrestrische Physik (MPE), Giessenbachstra{\ss}e 1, D-85748 Garching, Germany}

\author[0000-0002-5782-9093]{Daniel~A.~Dale}
\affiliation{Department of Physics and Astronomy, University of Wyoming, Laramie, WY 82071, USA}

\author[0000-0003-0946-6176]{Janice C. Lee}
\affiliation{Gemini Observatory/NSF’s NOIRLab, 950 N. Cherry Avenue, Tucson, AZ, USA}
\affiliation{Steward Observatory, University of Arizona, 933 N Cherry Ave,Tucson, AZ 85721, USA}

\author[0000-0002-5204-2259]{Erik Rosolowsky}
\affiliation{Department of Physics, University of Alberta, Edmonton, Alberta, T6G 2E1, Canada}

\author[0000-0002-3933-7677]{Eva Schinnerer}
\affiliation{Max-Planck-Institut f\"ur Astronomie, K\"onigstuhl 17, D-69117 Heidelberg, Germany}

\author[0000-0003-0410-4504]{Ashley.~T.~Barnes}
\affiliation{Argelander-Institut f\"{u}r Astronomie, Universit\"{a}t Bonn, Auf dem H\"{u}gel 71, 53121 Bonn, Germany}

\author[0000-0002-2545-5752]{Francesco Belfiore}
\affiliation{INAF — Arcetri Astrophysical Observatory, Largo E. Fermi 5, I-50125, Florence, Italy}

\author[0000-0003-0166-9745]{F. Bigiel}
\affiliation{Argelander-Institut f\"{u}r Astronomie, Universit\"{a}t Bonn, Auf dem H\"{u}gel 71, 53121 Bonn, Germany}

\author[0000-0002-5635-5180]{M\'elanie Chevance}
\affiliation{Universit\"{a}t Heidelberg, Zentrum f\"{u}r Astronomie, Institut f\"{u}r Theoretische Astrophysik, Albert-Ueberle-Stra{\ss}e 2, D-69120 Heidelberg, Germany}
\affiliation{Cosmic Origins Of Life (COOL) Research DAO, coolresearch.io}

\author[0000-0002-3247-5321]{Kathryn~Grasha}
\affiliation{Research School of Astronomy and Astrophysics, Australian National University, Canberra, ACT 2611, Australia}   
\affiliation{ARC Centre of Excellence for All Sky Astrophysics in 3 Dimensions (ASTRO 3D), Australia}   

\author[0000-0002-9768-0246]{Brent Groves}
\affiliation{International Centre for Radio Astronomy Research, University of Western Australia, 7 Fairway, Crawley, 6009 WA, Australia}

\author[0000-0002-8806-6308]{Hamid Hassani}
\affiliation{Department of Physics, University of Alberta, Edmonton, Alberta, T6G 2E1, Canada}

\author[0000-0002-9181-1161]{Annie~Hughes}
\affiliation{IRAP, Universit\'e de Toulouse, CNRS, CNES, UPS, (Toulouse), France}   

\author[0000-0002-0560-3172]{Ralf S.\ Klessen}
\affiliation{Universit\"{a}t Heidelberg, Zentrum f\"{u}r Astronomie, Institut f\"{u}r Theoretische Astrophysik, Albert-Ueberle-Stra{\ss}e 2, D-69120 Heidelberg, Germany}
\affiliation{Universit\"{a}t Heidelberg, Interdisziplin\"{a}res Zentrum f\"{u}r Wissenschaftliches Rechnen, Im Neuenheimer Feld 205, D-69120 Heidelberg, Germany}

\author[0000-0002-8804-0212]{J.~M.~Diederik~Kruijssen}
\affiliation{Cosmic Origins Of Life (COOL) Research DAO, coolresearch.io}

\author[0000-0003-3917-6460]{Kirsten L. Larson}
\affiliation{AURA for the European Space Agency (ESA), Space Telescope Science Institute, 3700 San Martin Drive, Baltimore, MD 21218, USA}

\author[0000-0001-9773-7479]{Daizhong Liu}
\affiliation{Max-Planck-Institut f\"{u}r Extraterrestrische Physik (MPE), Giessenbachstra{\ss}e 1, D-85748 Garching, Germany}

\author[0000-0002-1790-3148]{Laura A. Lopez}
\affiliation{Department of Astronomy, The Ohio State University, 140 West 18th Avenue, Columbus, OH 43210, USA}
\affiliation{Center for Cosmology and Astroparticle Physics, 191 West Woodruff Avenue, Columbus, OH 43210, USA}
\affiliation{Flatiron Institute, Center for Computational Astrophysics, NY 10010, USA}

\author[0000-0002-6118-4048]{Sharon E. Meidt}
\affil{Sterrenkundig Observatorium, Ghent University, Krijgslaan 281-S9, 9000 Gent, Belgium}

\author[0000-0001-7089-7325]{Eric J.\,Murphy}
\affiliation{National Radio Astronomy Observatory, 520 Edgemont Road, Charlottesville, VA 22903, USA}

\author[0000-0001-6113-6241]{Mattia C. Sormani}
\affiliation{Universit\"{a}t Heidelberg, Zentrum f\"{u}r Astronomie, Institut f\"{u}r Theoretische Astrophysik, Albert-Ueberle-Stra{\ss}e 2, D-69120 Heidelberg, Germany}

\author[0000-0002-8528-7340]{David A. Thilker}
\affiliation{Department of Physics and Astronomy, The Johns Hopkins University, Baltimore, MD 21218, USA}

\author[0000-0002-7365-5791]{Elizabeth~J.~Watkins}
\affiliation{Astronomisches Rechen-Institut, Zentrum f\"{u}r Astronomie der Universit\"{a}t Heidelberg, M\"{o}nchhofstra\ss e 12-14, 69120 Heidelberg, Germany}
\suppressAffiliations

\begin{abstract}

We present maps of the 3.3 \micron\ polycyclic aromatic hydrocarbon (PAH) emission feature in NGC~628, NGC~1365, and NGC~7496 as observed with the Near-Infrared Camera (NIRCam) imager on JWST from the PHANGS-JWST Cycle~1 Treasury project. We create maps that isolate the 3.3~\micron\ PAH feature in the F335M filter (F335M$_{\rm PAH}$) using combinations of the F300M and F360M filters for removal of starlight continuum. This continuum removal is complicated by contamination of the F360M by PAH emission and variations in the stellar spectral energy distribution slopes between 3.0 and 3.6~\micron. We modify the empirical prescription from \citet{Lai2020} to remove the starlight continuum in our highly resolved galaxies, which have a range of starlight- and PAH-dominated lines-of-sight.  Analyzing radially binned profiles of the F335M$_{\rm PAH}$ emission, we find that between $5-65$\% of the F335M intensity comes from the 3.3~\micron\ feature within the inner 0.5 $r_{25}$ of our targets. This percentage systematically varies from galaxy to galaxy, and shows radial trends within the galaxies related to each galaxy's distribution of stellar mass, interstellar medium, and star formation. The 3.3~\micron\ emission is well correlated with the 11.3~\micron\ PAH feature traced with the MIRI F1130W filter, as is expected, since both features arise from C-H vibrational modes. The average F335M$_{\rm PAH}$/F1130W ratio agrees with the predictions of recent models by \citet{Draine2021} for PAHs with size and charge distributions shifted towards larger grains with normal or higher ionization.

\end{abstract}


\section{Introduction} \label{sec:intro}

Polycyclic aromatic hydrocarbon (PAH) emission is observed ubiquitously in the interstellar medium (ISM) of most massive, star-forming galaxies, carrying 10-20\% of the total infrared emission \citep{Draine2007b,smith2007,tielens2008,li2020}. Consequently, these PAH vibrational bands are detectable in galaxies out to high redshift \citep[e.g.][]{riechers2014}. In the era of {\em JWST}, PAH emission will be a widely observed and valuable tracer of the ISM across a large range of redshifts. In particular, 3.3 \micron\ PAH emission -- the shortest wavelength PAH feature of significant strength -- will potentially be detectable in galaxies out to $z\sim7$ with JWST MIRI spectroscopy.

The 3.3 \micron\ PAH feature arises from a C-H stretching vibration \citep{schutte1993,vandiedenhoven2004}.  Laboratory and theoretical studies suggest that the feature is dominated by small, neutral PAHs \citep{maragkoudakis2020,Draine2021,kerkeni2022}. The longer wavelength C-H bending mode at 11.3 \micron\ is also dominated by neutral PAHs but is expected to arise from larger grains, and the ratio of 3.3 \micron\ to 11.3 \micron\ is therefore expected to be a relatively clean diagnostic of the average size of the PAHs \citep{maragkoudakis2020,Lai2020,Draine2021}. 

\begin{figure}
    \centering
    \includegraphics[width=0.5\textwidth]{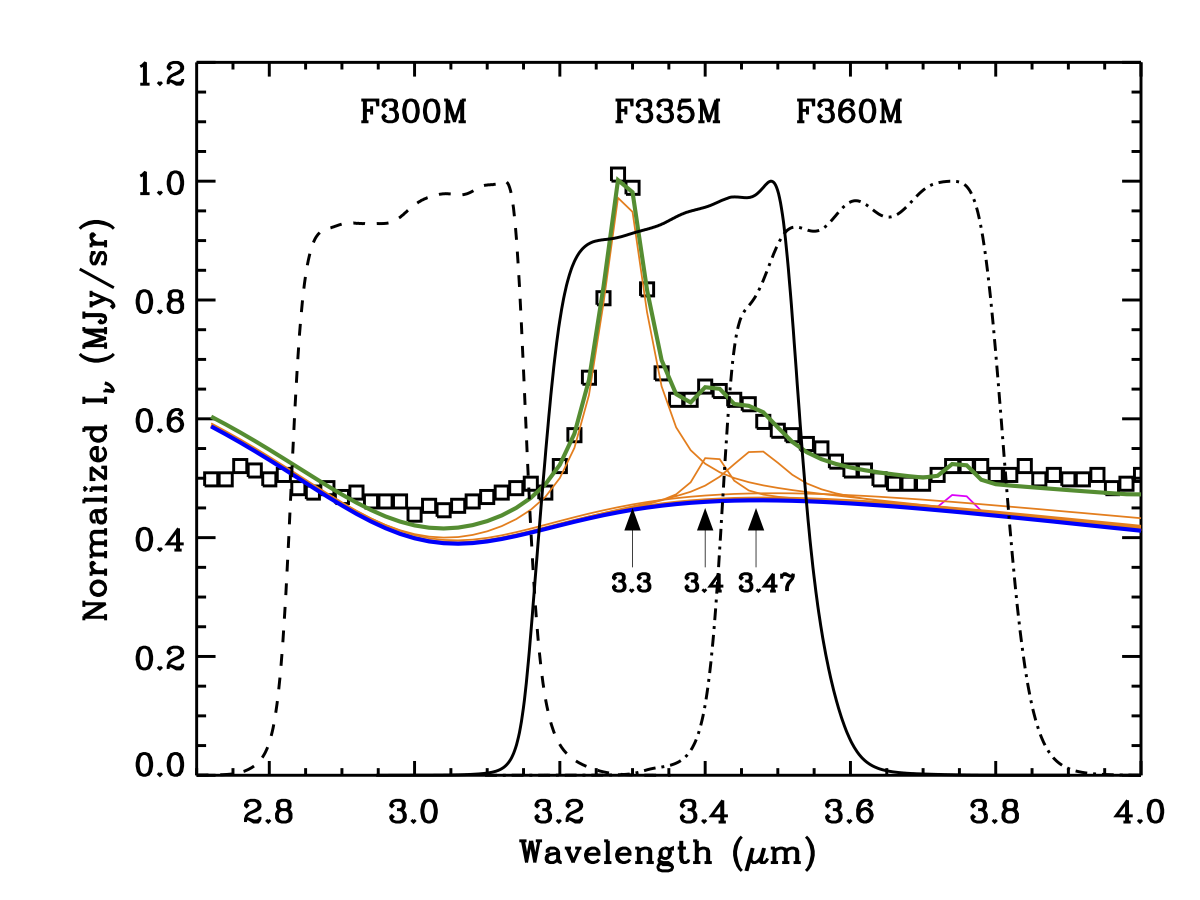}
    \caption{The NIRCam F300M, F335M, and F360M wavelength coverage overlaid on the 1C template spectrum from \citet{Lai2020} (square points) with their PAHFIT-based decomposition \citep{smith2007}. The continuum (including starlight, hot dust, and the effects of attenuation) is shown in blue. Orange lines show the individual dust features, including the 3.3 \micron\ PAH feature, the 3.4 \micron\ ``aliphatic'' emission feature and the 3.47 \micron\ plateau, which are labeled with arrows. The 3.74 \micron\ Pfund $\gamma$ emission line is shown in magenta.  The combined fit too all components is shown in green. PAH (or aliphatic) dust emission features contribute both to F335M and F360M, as well as longer wavelength dust emission in the F360M filter.  These contributions complicate continuum subtraction in our highly resolved galaxies, where individual lines of sight may be dominated by PAH emission in all three filters.}
    \label{fig:spec33}
\end{figure}

The 3.3 \micron\ PAH feature was not covered by the spectroscopic instruments on Spitzer. Due to the wide wavelength coverage of the Spitzer IRAC 3.6 \micron\ filter, it was difficult to isolate the 3.3 \micron\ feature contribution to the broadband emission \citep[though principal component approaches have found a signal correlated with longer wavelength PAH emission;][]{meidt2012,querejeta2015}. Observations with ISO and Akari measured the 3.3 \micron\ feature spectroscopically \citep[e.g.][]{verstraete2001,imanishi2010,lee2012,Lai2020}, as have some ground-based studies \citep[e.g.][]{sloan1997}, but these efforts have generally focused on very bright targets---Milky Way photodissociation regions, ultraluminous infrared galaxies (ULIRGs), and galaxy nuclei, for example.  JWST observations provide the first opportunity to map 3.3 \micron\ emission  in nearby galaxies, including both star-forming regions and the diffuse ISM.  Moreover, the NIRCam medium-band filter set of F300M, F335M, and F360M (see Figure~\ref{fig:spec33} for an illustration) provides one of the first opportunities to make large area maps of 3.3 \micron\ emission, and use the comparison to 11.3 \micron\ PAH emission traced by the MIRI F1130W filter to map PAH size variations.  The NIRCam point spread function at F300M, F335M, and F360M has FWHM $\sim0.10-0.12''$, yielding $5-10$ pc resolution in galaxies at distances of $10-20$ Mpc.

There are some challenges in isolating the 3.3 \micron\ PAH feature using the NIRCam medium bands. The variation of the stellar spectral energy distribution appears to be large enough that scaling F300M with a single stellar color does not provide a clean subtraction (see Section~\ref{sec:sub}).  While using a linear interpolation between F300M and F360M can address this issue and remove stellar continuum with a varying slope, contamination of the F360M band by the 3.3 \micron\ feature itself and the nearby 3.4 \micron\ ``aliphatic'' feature \citep[thought to arise from PAHs with an aliphatic sub-group, which encompasses a variety of non-aromatic hydrocarbon structures that may be attached to the PAHs;][]{yang2017} and the 3.47 \micron\ ``plateau'' features also attributed to PAHs \citep{hammonds2015,Lai2020} complicates the linear interpolation in places with bright PAH emission.  Both features are substantially weaker than the 3.3 \micron\ PAH \citep[for instance, the 3.4 \micron\ feature is on average $\sim8$\% of the strength of 3.3 \micron\ feature;][]{Lai2020}, but they fall in the F360M band, potentially complicating the continuum subtraction.  The degree to which the 3.3 and 3.4 \micron\ features vary relative to each other is not yet well characterized in normal star-forming galaxies \citep[though Akari observations have shown variations in starbursts and luminous infrared galaxies;][]{Lai2020}. Spectroscopic observations with NIRSpec in nearby galaxies are well suited to addressing this question in the near future. The 3.3 \micron\ feature itself can significantly contribute to both of the nearby medium bands in cases where the PAH emission is bright compared to the underlying continuum.  For nearby galaxies, where individual stars and diffuse emission are highly resolved, we expect to find regions with different relative contributions from PAHs and starlight, making it necessary to carefully test any continuum subtraction procedure in both regimes.

In this Letter, we describe the production of the first 3.3~\micron\ PAH maps of NGC~628, NGC~1365, and NGC~7496 from the PHANGS survey.  Our goal is to demonstrate this key capability of JWST to map the ISM at high angular resolution ($\sim0.1''$) with the NIRCam medium bands. The PHANGS-JWST Treasury program will eventually cover 19 galaxies from the PHANGS sample (see \citealt{LEE_PHANGSJWST} for more details). These targets have deep, high-resolution, ancillary information from ALMA \citep{leroy2021}, VLT-MUSE \citep{emsellem2022}, Hubble \citep{lee2022}, AstroSat (Hassani et al.\ in prep), and more. Combining NIRCam imaging of the 3.3 \micron\ PAH feature with MIRI imaging of the 7.7 and 11.3 \micron\ features enables one of the first studies of the variation of both PAH size and charge over large regions of nearby galaxies \citep{CHASTENET2_PHANGSJWST}, in H\textsc{ii} regions \citep{EGOROV_PHANGSJWST}, and near young star clusters and associations \citep{DALE_PHANGSJWST,RODRIGUEZ_PHANGSJWST}.   As part of this effort we found it was necessary to adjust the current best NIRCam medium band continuum removal prescription in the literature, from \citet{Lai2020}, to account for PAH emission (or related dust emission) contamination of F360M filter (Section~\ref{sec:sub}). While we expect the optimal recipe may evolve, especially informed by future JWST NIRSpec spectral mapping, our results already demonstrate that the NIRCam medium band filter set is one of the most powerful available tools to map the ISM at high resolution and offers a new window into PAH properties. Our work here also highlights the need for future spectroscopic calibration of the empirical medium-band PAH continuum subtraction recipes.   

\section{Observations} \label{sec:obs}

\begin{deluxetable}{lcccc}[t!]
\tabletypesize{\small}
\tablecaption{Galaxy Properties \label{tab:galaxies}}
\tablewidth{0pt}
\tablehead{
\colhead{Target} & 
\colhead{Distance} &
\colhead{Inclination} &
\colhead{P.A.} &
\colhead{r$_{25}$} \\
\colhead{} & 
\colhead{(Mpc)} &
\colhead{($^{\circ}$)} &
\colhead{($^{\circ}$)} &
\colhead{(kpc)}
}
\startdata
NGC~628 &  9.8 & 9 & 21 & 14.1 \\  
NGC~1365 &  19.6 & 55 & 201 & 34.2 \\
NGC~7496 &  18.7 & 36 & 194 & 9.1
\enddata
\tablecomments{Properties adopted from \citet{leroy2021} following \citet{LEE_PHANGSJWST}, which draws orientations from \citet{lang20} and distances from \citet{anand21}.}
\end{deluxetable}

We use observations of the first three galaxies from the PHANGS-JWST survey \citep{LEE_PHANGSJWST} observed with NIRCam---NGC~628, NGC~1365, and NGC~7496.  The properties we adopt for the galaxies are listed in Table~\ref{tab:galaxies}.
The three galaxies were observed with the long wavelength (LW) channel of the NIRCam instrument on JWST with the F300M, F335M, and F360M filters between July and August 2022. The targets were covered with small 2- or 4-pointing mosaics. The total integration for each mosaic tile was 386.5 seconds in the F300M and F335M filters and 429.4 seconds in F360M.  The resulting uncertainties in the images are $\sim$0.05, 0.045, and 0.06 \mjysr\ for the F300M, F335M, and F360M bands. These observations were obtained simultaneously with deep F200W observations in the SW (short wavelength) channel. After the NIRCam observations, a similar region of the galaxy was observed with the MIRI instrument, using the F770W, F1000W, F1130W, and F2100W filters. The observations and data processing are described in detail in \citeauthor{LEE_PHANGSJWST} in this Issue. The data processing included a step of correcting for astrometric offsets between the NIRCam images. In the following analysis, in addition to the F300M, F335M, and F360M observations, we also make use of the F200W and F1130W maps\footnote{The filter widths of the F200W, F300M, F335M, F360M, F1130W filters are $\Delta \micron =$ 0.461, 0.318, 0.347, 0.372, and 0.7 \micron.}. F200W primarily samples stellar continuum, providing a template for stellar emission that is relatively uncontaminated by strong PAH or hot dust emission.  The F1130W filter, on the other hand, is quite narrowly centered on the 11.3 \micron\ PAH feature, yielding a measurement that in essentially all cases is dominated by PAH emission. The contribution of starlight at 11.3 \micron\ is minimal and over the 0.7 \micron\ wavelength filter width of F1130W, PAH emission greatly exceeds the contribution from other small, stochastically heated dust grains. Spectroscopy from the ``Spitzer Infrared Nearby Galaxies Survey'' (SINGS) shows that wavelengths around 11.3 \micron\  are dominated by PAH emission in $\sim$Z$_{\odot}$ galaxies like our targets \citep{smith2007, whitcomb2022}.

To correctly trace the faint diffuse emission in the images, we made small adjustments to the background level of the images ($\sim$0.1 \mjysr). In the pipeline-produced data products, the NIRCam maps have a small but significant negative offset that is evident in faint regions of the map.  In each of the galaxies, we found approximately empty sky regions outside the galaxy to determine the background level. We measured the average values in these regions using a biweight mean algorithm to reject outliers. We found the average background level from all empty sky regions and subtracted it from the map.   After adding back in the offsets, all maps reach $\sim0$ \mjysr\ in their outskirts, without becoming significantly negative. The origin of the background offsets is not yet known and may be resolved with future updates to the NIRCam processing and calibration pipeline.

To investigate the optimal continuum subtraction for the 3.3 \micron\ feature, in Section~\ref{sec:sub} we explore use of the F300M and F360M filters to predict the continuum in the F335M image.   
For this comparison, we do not convolve the images to matched resolution at this time, since the small differences in FWHM mean that kernel generation is subject to artifacts and noise amplification. Our tests indicate that the FWHM of the PSF is similar across these three bands to within $\sim10\%$. 
We use the F1130W to select regions with strong PAH emission in the images.  We do not convolve to match F1130W since it is used only to select coarse regions of the image (i.e. F1130W $<1$ or $>10$ \mjysr).
We regrid the maps to the common astrometric grid of the F335M data with pixel scale $0.063''$ using bilinear interpolation.

We note that the PHANGS-JWST NIRCam imaging is affected by $1/f$ noise. As described in \citet{LEE_PHANGSJWST}, we have applied a correction that minimizes the striping.  However, in the difference images between F300M, F335M, and F360M used in this Letter, the stripes are prominent even after correction. Future work or updates to the calibration files and pipeline processing may be able to remove this noise.

\section{Subtracting 3.3 \micron\ Continuum} \label{sec:sub}

\begin{figure*}[ht]
    \centering
    \includegraphics[width=\textwidth]{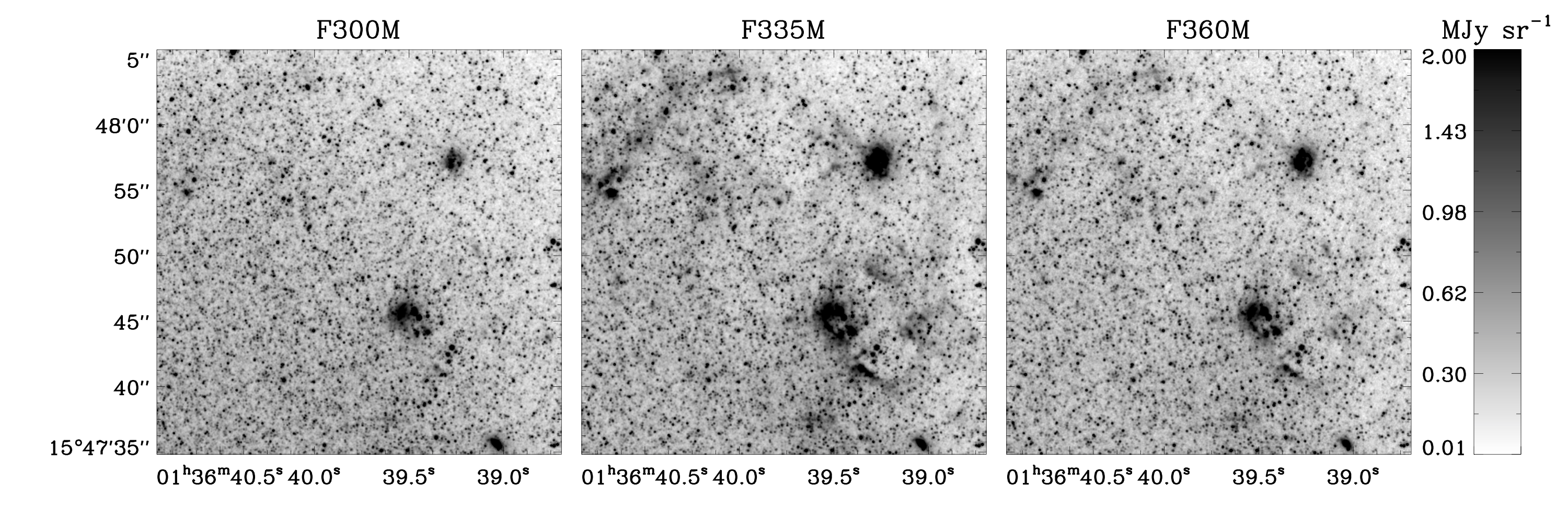}
    \includegraphics[width=\textwidth]{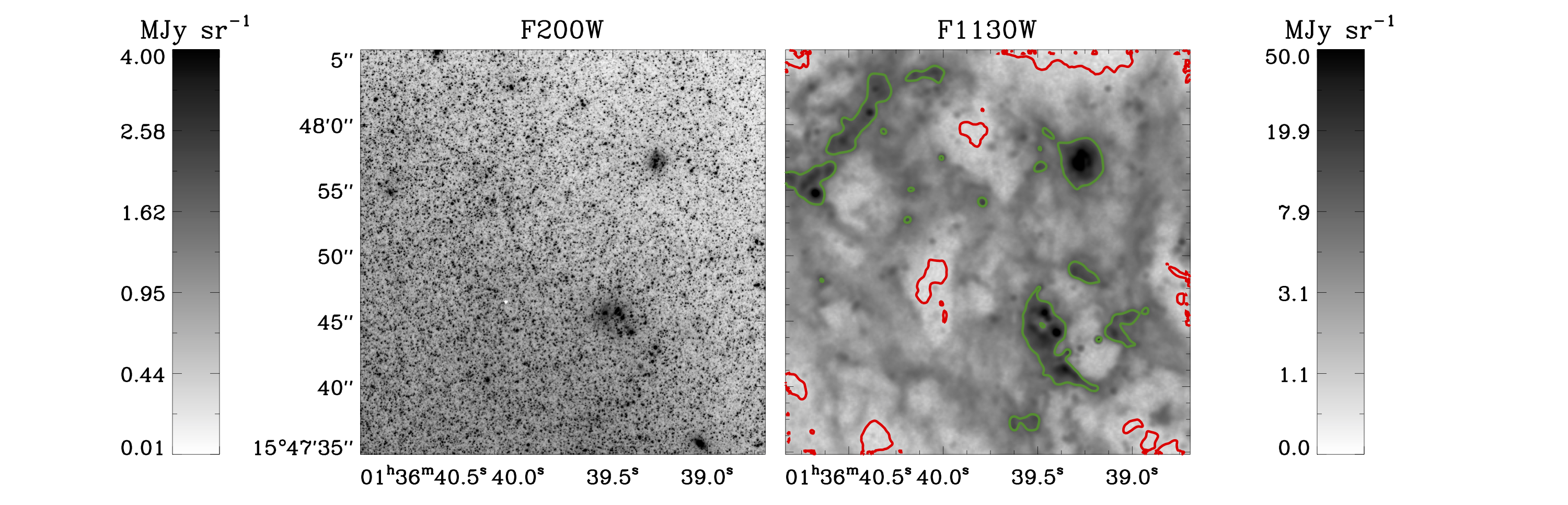}
    \caption{A representative $1.5\times 1.5$ kpc region of NGC~628 shown in the F200W, F300M, F335M, F360M, and F1130W bands with an {\em asinh} color table at each filter's native resolution. The F1130W filter traces primarily PAH emission from the 11.3 \micron\ feature. The F200W filter traces primarily stellar continuum.  The F335M filter is centered on the 3.3 \micron\ PAH feature and includes both stellar continuum and PAH emission.  Diffuse emission is visible in the F335M and F360M bands. In the F1130W panel, we highlight faint PAH emission with F1130W $<1$ \mjysr\ with a red contour and bright PAH emission with F1130W $>10$ \mjysr\ with a green contour.  These faint and bright selections are used in Figure~\ref{fig:colorvs113}.}
    \label{fig:ngc628zoom}
\end{figure*}

In Figure~\ref{fig:ngc628zoom} we show images of a representative $1.5\times1.5$ kpc region in NGC~628 in all the photometric bands that we consider in this analysis.  We will use this region to visualize the results of our continuum subtraction. The F300M, F335M, and F360M images are shown with identical intensity scaling.  Diffuse emission can be seen clearly in the F335M and F360M filters. In bright star-forming regions, diffuse emission is also evident in F300M and F200W, potentially arising from hot dust emission and/or nebular emission lines covered by those filters. The diffuse emission present in the F360M filter could be the result of a combination of 3.3 \micron\ emission, other nearby emission lines, hot dust continuum, and/or the 3.4 \micron\ aliphatic feature and faint 3.47 \micron\ PAH ``plateau'' feature. 

\subsection{\citet{Lai2020} Prescription for 3.3 \micron\ Continuum Subtraction}

Previous work by \citet{Lai2020} using a combination of Akari and Spitzer spectroscopy determined a set of coefficients for a linear combination of the F300M and F360M bands to remove the continuum, following the form: 
\begin{equation}
    {\rm F335M}_{\rm PAH} = {\rm F335M} -  {\rm F335M}_{\rm cont}.
\end{equation}
\begin{equation}
    {\rm F335M}_{\rm cont} = A \times {\rm F300M} + B \times {\rm F360M}
\end{equation}
Their recommended coefficients are $A_{\rm Lai}=0.35$ and $B_{\rm Lai}=0.65$. 
For the current investigation, we work in surface brightness units (\mjysr) and do not integrate over the filter width or convert to $\nu F_{\nu}$. Therefore, our F335M$_{\rm PAH}$ and F335M$_{\rm cont}$ maps are in units of \mjysr.  Following \cite{Lai2020}, F335M$_{\rm PAH}$ can be converted to an integrated 3.3 \micron\ band flux in units of $10^{-14}$ erg s$^{-1}$ cm$^{-2}$ by multiplying with a factor of 10.78 (appropriate for nearby galaxies at $z=0$). The \citet{Lai2020} prescription predicts colors for the F335M continuum described by:
\begin{equation}\label{eq:laicolor}
    \frac{{\rm F335M}_{\rm cont}}{\rm F300M} = A_{\rm Lai} + B_{\rm Lai} \frac{\rm F360M}{\rm F300M}
\end{equation}

This calibration was derived for moderately obscured galaxies with high star formation rates. \citet{Lai2020} find through synthetic photometry that their targets are typically continuum dominated in the F335M band, with $\sim$15\% of the flux being due to the PAH feature. \citet{Lai2020} note that water ice absorption at 3.05 \micron\ in the F300M band may influence their results for the more heavily obscured targets.  Our sample has minimal obscuration on average (median E(B$-$V) $\sim0.2-0.3$ for HII regions in these targets; Groves et al.\ submitted) and lower star formation surface densities than most of the \citet{Lai2020} sample.  In the following, we investigate continuum subtraction with the three medium bands, but we note that unlike \citet{Lai2020} we are unable to spectroscopically constrain any contamination from the 3.4 \micron\ ``aliphatic'' or 3.47 \micron\ ``plateau'' features.  These features are significantly fainter than the 3.3 \micron\ band, so may not dominate the diffuse signal in F360M.

\subsection{Observed F300M-F335M-F360M Colors in PAH- and Continuum-Dominated Regions}

The PHANGS-JWST observations resolve individual stars and ISM emission at $5-40$ pc scales in the F300M to F1130W bands.  From Figure~\ref{fig:ngc628zoom}, it is clear that there is a wide variation in the degree to which any given line of sight is continuum or PAH dominated. In addition, it is clear that in places where the emission is PAH dominated, the F360M band traces PAH related emission as well.  This suggests that a simple linear interpolation in the continuum following Equation~\ref{eq:laicolor} will not suffice.  

\begin{figure*}
    \centering
    \includegraphics[width=0.85\textwidth]{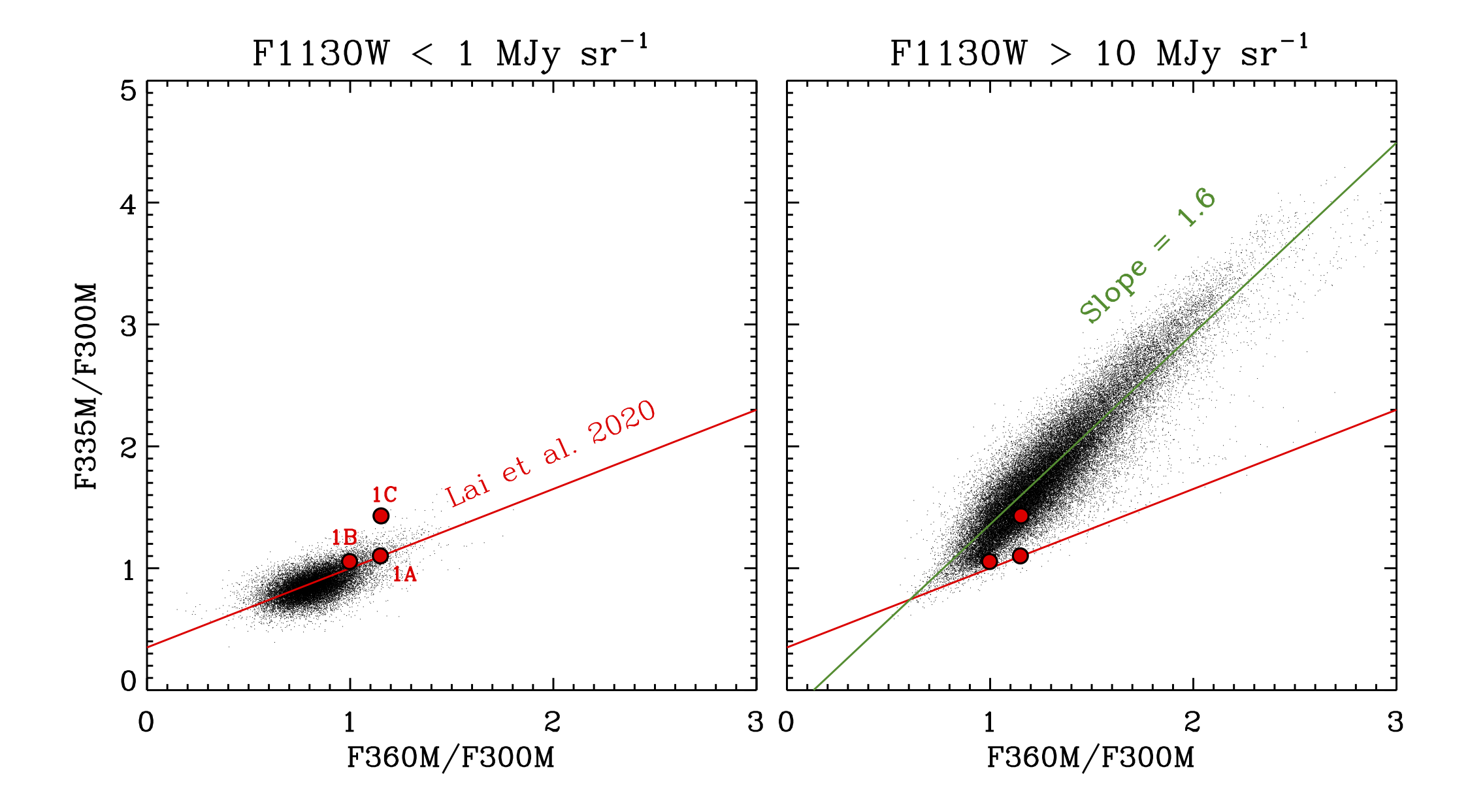}
    \caption{F335M/F300M versus F360M/F300M color in our representative region of NGC~628 selected by F1130W surface brightness. On the left, the PAH emission at F1130W is faint (F1130W $<1$ \mjysr), so the 3.0-3.6 \micron\ colors should be dominated by stars. This region is highlighted with a red contour in Figure~\ref{fig:ngc628zoom}. On the right, we select bright regions in F1130W ($>10$ \mjysr), for which we expect the colors to be dominated by PAH emission. The PAH-bright region is highlighted with a green contour in Figure~\ref{fig:ngc628zoom}. The red line in each panel shows the \citet{Lai2020} prescription for the stellar continuum. In starlight dominated regions, this prescription does well at predicting the continuum at F335M.  However, in regions dominated by PAHs, the F360M/F300M color is also responding to the PAH emission, leading to overestimates of the F335M continuum. We include synthetic photometry for the 1A, 1B, and 1C template spectra from \citet{Lai2020} with red symbols, illustrating that our observed colors span a much wider range than the spectra used to create the continuum subtraction prescription.}
    \label{fig:colorvs113}
\end{figure*}

To investigate these issues, we measure the F300M/F335M/F360M colors in regions of the images we expect to be continuum dominated versus PAH dominated using F1130W as a guide.  We first select all regions where F1130W $< 1$ \mjysr\ to represent a low PAH emission region (shown with a red contour in the bottom right panel of Figure~\ref{fig:ngc628zoom}).  We also select regions where F1130W $>10$ \mjysr, representing highly PAH emission dominated regions (green contour in bottom right panel of Figure~\ref{fig:ngc628zoom}).  In Figure~\ref{fig:colorvs113}, we show the results of this selection for pixels above a 5$\sigma$ detection threshold in the F300M, F335M, and F360M bands. 

On Figure~\ref{fig:colorvs113} we also include synthetic photometry in the NIRCam medium bands for the 1A, 1B, and 1C template spectra from \citet{Lai2020}. These template spectra show only moderate attenuation, making them the most comparable templates to our targets. The template spectra, which represent the spectra used to formulate the \citet{Lai2020} continuum prescription cover a much narrower range of F360M/F300M colors than our observations.

For the low PAH emission regions (F1130W $<1$ \mjysr), we find $3.0-3.6$ \micron\ colors in good agreement with the \citet{Lai2020} prediction.  These measurements also suggest that starlight continuum at these wavelengths is not well described by a single color. This can be seen in the extension of the both the F335M/F300M and F360M/F300M colors along the \citet{Lai2020} slope.  If a single stellar color would suffice, these points would be expected to be clustered around a single value of F335M/F300M and F360M/F300M. The range of colors suggests we are seeing contributions from a variety of stellar populations, including red supergiants and asymptotic giant branch stars which show a wider range of near- to mid-IR spectral shapes, some related to circumstellar dust \citep{meidt2012}.   

From the strongly PAH-dominated regions identified by high F1130W surface brightness (F1130W $>10$ \mjysr) in Figure~\ref{fig:colorvs113}, we see that PAH emission appears in the color-color diagram with a linear slope $B_{\rm PAH} = 1.6$ and offset $A_{\rm PAH}=-0.2$, where we describe the linear relation in the colors with:
\begin{equation}
    \frac{{\rm F335M}_{\rm PAH}}{\rm F300M} = A_{\rm PAH} + B_{\rm PAH} \frac{\rm F360M}{\rm F300M}
\end{equation}
We found the same slope using all pixels in the three target galaxies where F1130W $>10$\mjysr, suggesting that the $B_{\rm PAH}=1.6$ slope is a good representation of the colors of PAH dominated emission across our sample.
This comparison shows that in regions dominated by PAH emission, the F360M/F300M color and the F335M/F300M color are both tracing the PAH feature(s) in this wavelength range. Performing a linear interpolation with the \citet{Lai2020} slope would lead to subtracing off 3.3 \micron\ PAH emission, because the F360M filter is capturing emission from this feature.

\subsection{An Optimized Continuum Subtraction Recipe for Highly Resolved Galaxies}

\begin{figure}
    \centering
    \includegraphics[width=0.45\textwidth]{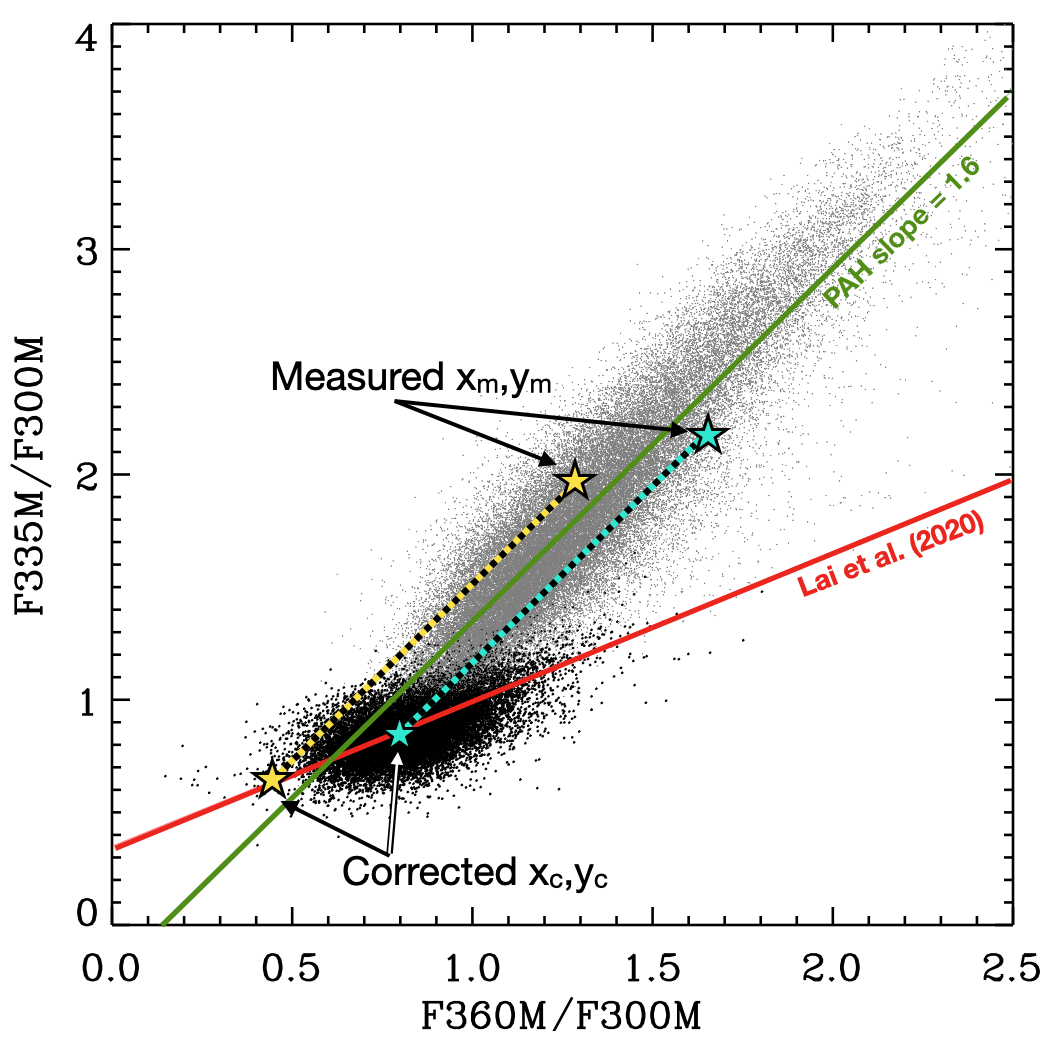}
    \caption{Illustration of the correction to the F360M/F300M color to minimize over-subtraction of PAH emission from the F335M filter. The correction works by scaling the colors along the PAH-dominated color trend which has a slope of $B_{\rm PAH} = 1.6$. The black points show the pixel-by-pixel colors derived in PAH-faint regions of the map (F1130W $<1$ \mjysr). The gray points show colors in the PAH-bright regions (F1130W $>10$ \mjysr). Two examples of the correction are shown with yellow and cyan stars. The measured values $x_m, y_m$ are scaled along the PAH slope till it intersects the \citet{Lai2020} relation for stellar continuum colors to obtain corrected values $x_c, y_c$. }
    \label{fig:corrdiagram}
\end{figure}

To avoid oversubtraction of PAH emission, we determine a correction for the predicted F335M$_{\rm cont}$ based on the observed F335M/F300M and F360M/F300M colors. This approach represents a first order correction to the continuum subtraction approach from \citet{Lai2020}, and future effort on spectroscopic 3.3 \micron\ PAH observations will be necessary to develop a more rigorous procedure for highly resolved targets like ours.  The nature of this correction is shown in Figure~\ref{fig:corrdiagram}. 
The basic approach is to use the observed F335M/F300M color as an indication of how PAH-dominated the emission is in a given location.  Using the observed slope of $B_{\rm PAH} = 1.6$ for PAH-dominated emission, we can scale the F360M/F300M color to where it intersects the relationship from \citet{Lai2020}, obtaining a corrected F360M/F300M color with which to predict the F335M$_{\rm cont}$. The intersection of these two lines is where the PAH contribution to both colors is assumed to be zero.

For each point, we scale the measured F360M/F300M value ($x_m$) and F335M/F300M value ($y_m$) along the PAH slope to where it intersects with the continuum relationship described by \citet{Lai2020}.  This yields a corrected F360M/F300M ratio ($x_c$) and F335M/F300M ratio ($y_c$). The PAH slope is described as:
\begin{equation}\label{eq:pahslope}
    y = A_{\rm PAH} + B_{\rm PAH} x,
\end{equation}
and the \citet{Lai2020} slope is given by:
\begin{equation}\label{eq:laislope}
    y = A_{\rm Lai} + B_{\rm Lai} x.
\end{equation}
Using Equation~\ref{eq:pahslope}, we can write the relationship between $x_m, y_m$ and $x_c, y_c$ as:
\begin{equation}
    \frac{y_m - y_c}{x_m - x_c} = B_{\rm PAH}.
\end{equation}
The corrected values will lie on Equation~\ref{eq:laislope}, so:
\begin{equation}
    y_c = A_{\rm Lai} + B_{\rm Lai} x_c
\end{equation}
We can then write $x_c$ in terms of the measured colors ($x_m, y_m$) as follows:
\begin{equation}
    x_c = \frac{B_{\rm PAH} x_m - y_m + A_{\rm Lai}}{B_{\rm PAH} - B_{\rm Lai}}.
\end{equation}
Putting this back into the \citet{Lai2020} formula, we can then obtain a prediction for $y_c$, which represents the F335M/F300M color appropriate for continuum, and subsequently the F335M continuum as follows:
\begin{equation}\label{eq:yc}
    y_c = B_{\rm Lai} \left( \frac{B_{\rm PAH} x_m - y_m + A_{\rm Lai}}{B_{\rm PAH} - B_{\rm Lai}} \right) + A_{\rm Lai}
\end{equation}
\begin{equation}\label{eq:finalcont}
    {\rm F335M}_{\rm cont}  = y_c \times {\rm F300M} .
\end{equation}

\begin{figure*}
    \centering
    \includegraphics[width=\textwidth]{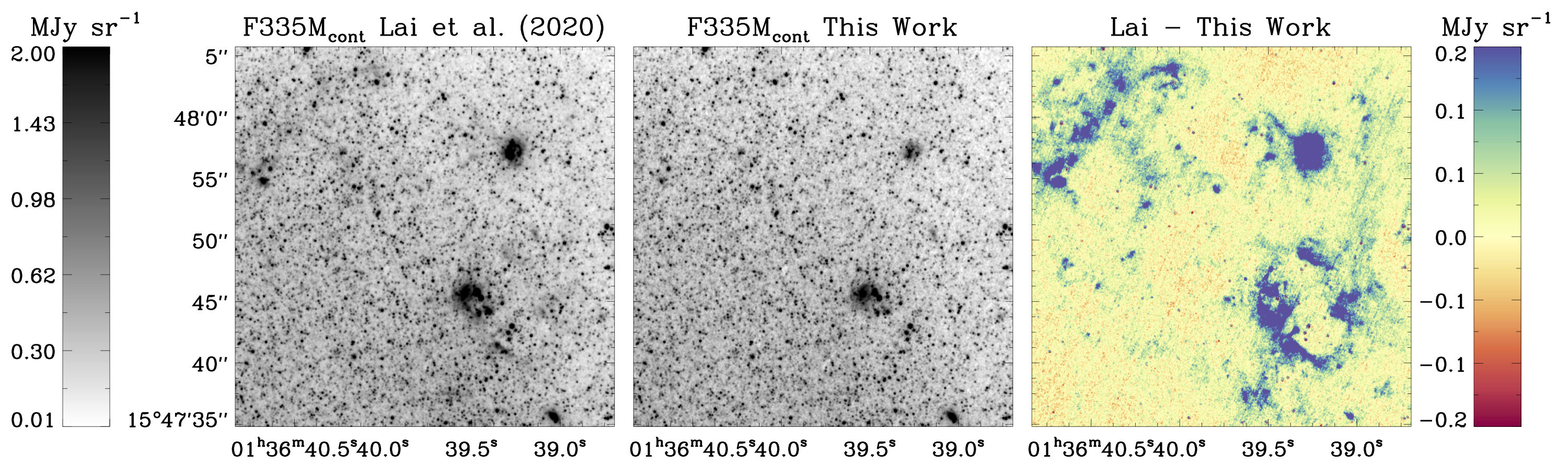}
    \caption{The predicted F335M$_{\rm cont}$ from the \citet{Lai2020} formula in Equation~\ref{eq:laicolor} (left), the empirical prescription presented in this work in Equations~\ref{eq:yc} and ~\ref{eq:finalcont} (middle), and the difference between them (right) for our representative region from NGC~628 shown in Figure~\ref{fig:ngc628zoom}. The left and middle panels show the same 1.5$\times$1.5 kpc region of NGC~628 shown in Figure~\ref{fig:ngc628zoom} with the same $asinh$ color table. The right panel shows that significant amounts of diffuse emission are included in the F335M$_{\rm cont}$ from the \citet{Lai2020} prescription because of the fact that F360M also includes some PAH-related emission. Our prescription minimizes this contamination by correcting the F360M/F300M colors while still cleanly subtracting the stellar continuum.}
    \label{fig:mysub}
\end{figure*}

In Figure~\ref{fig:mysub} we show a comparison of the F335M continuum derived from the above method and that from \cite{Lai2020}. The right panel shows the difference in the continuum predicted by the two techniques. It is clear that the corrected formula more cleanly isolates stellar continuum, minimizing over-subtraction of PAH emission, while maintaining the success of the \citet{Lai2020} formula at subtracting starlight.  While this empirical correction can be fine-tuned in the future, this approach provides a straightforward technique to measure the 3.3 \micron\ feature for these first science applications with NIRCam medium band imaging.  We proceed in the following Section to interpret the resulting maps of PAH emission.

In the future, we plan to investigate more sophisticated approaches to determining the PAH contamination of the various NIRCam medium bands. The basic assumptions in this approach are that the F300M is relatively uncontaminated by diffuse emission, and the F335M/F300M ratio therefore gives a reasonable estimation of the degree of PAH emission. However, since there is a range of colors appropriate for the stars, a single scaled stellar spectral energy distribution tied to F300M does not appear to work well at removing the continuum.  Instead, we interpret the spread in F360M/F300M at a fixed F335M/F300M as variations in the stellar SED.  This may not be the case, as various other effects can alter the 3.0-3.6 \micron\ colors, including variations in the 3.4 \micron\ feature, extinction, ice absorption features, other diffuse emission from hot dust and/or nebular emission. Future work comparing the $3.0-3.6$ \micron\ stellar colors with stellar population modeling based on the PHANGS-MUSE observations \citep{emsellem2022} may allow development of more precise stellar continuum recipes.  Future spectroscopic calibration of F335M continuum subtraction recipes will be critical to fully exploit the capability to map PAH emission with medium band filters on NIRCam and move beyond the first-order correction presented here.

\section{Results}\label{sec:results}

\begin{figure*}
    \centering
    \includegraphics[width=\textwidth]{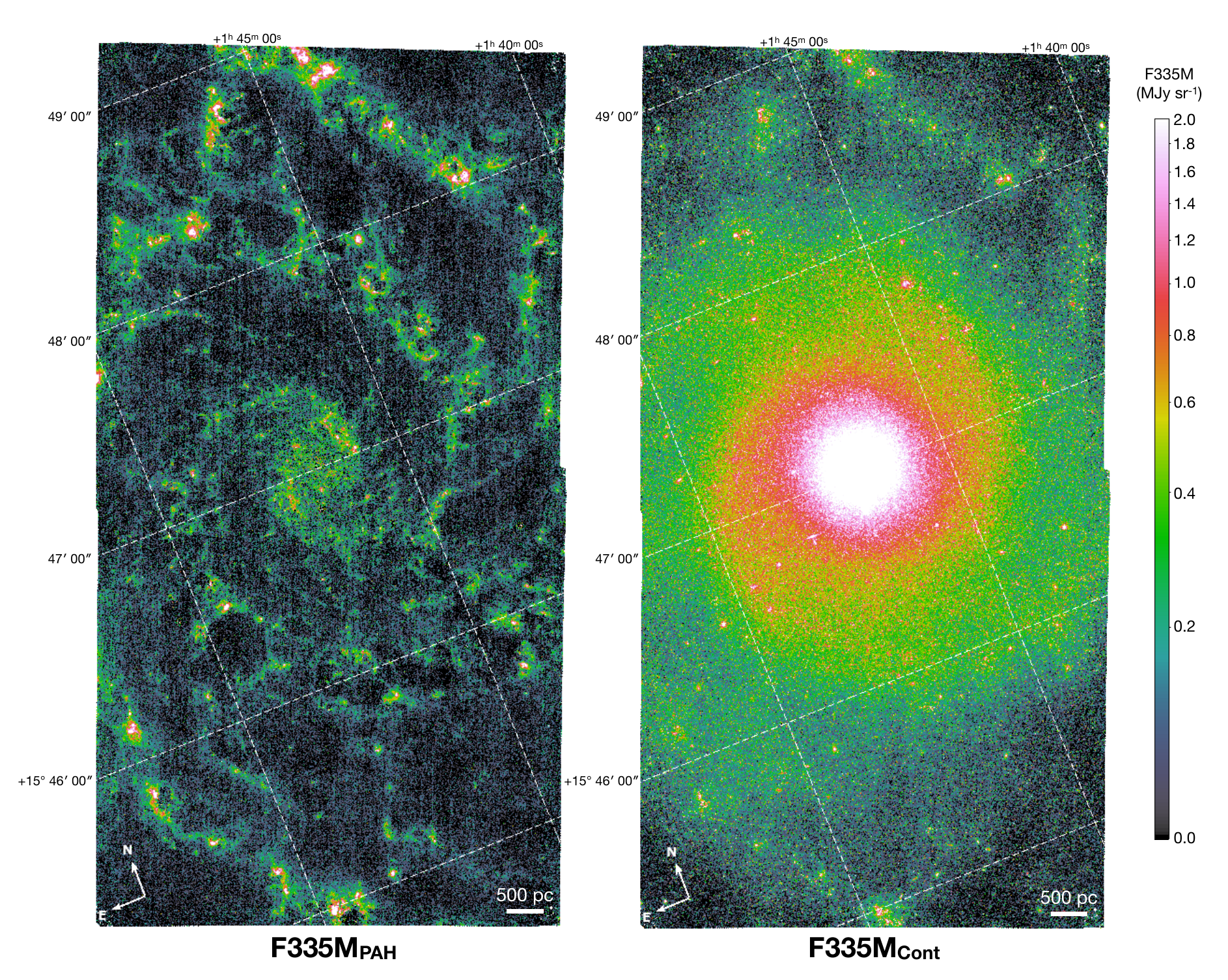}
    \caption{PAH 3.3 \micron\ emission and continuum for NGC~628.}
    \label{fig:map_ngc628}
\end{figure*}

\begin{figure*}
    \centering
    \includegraphics[width=\textwidth]{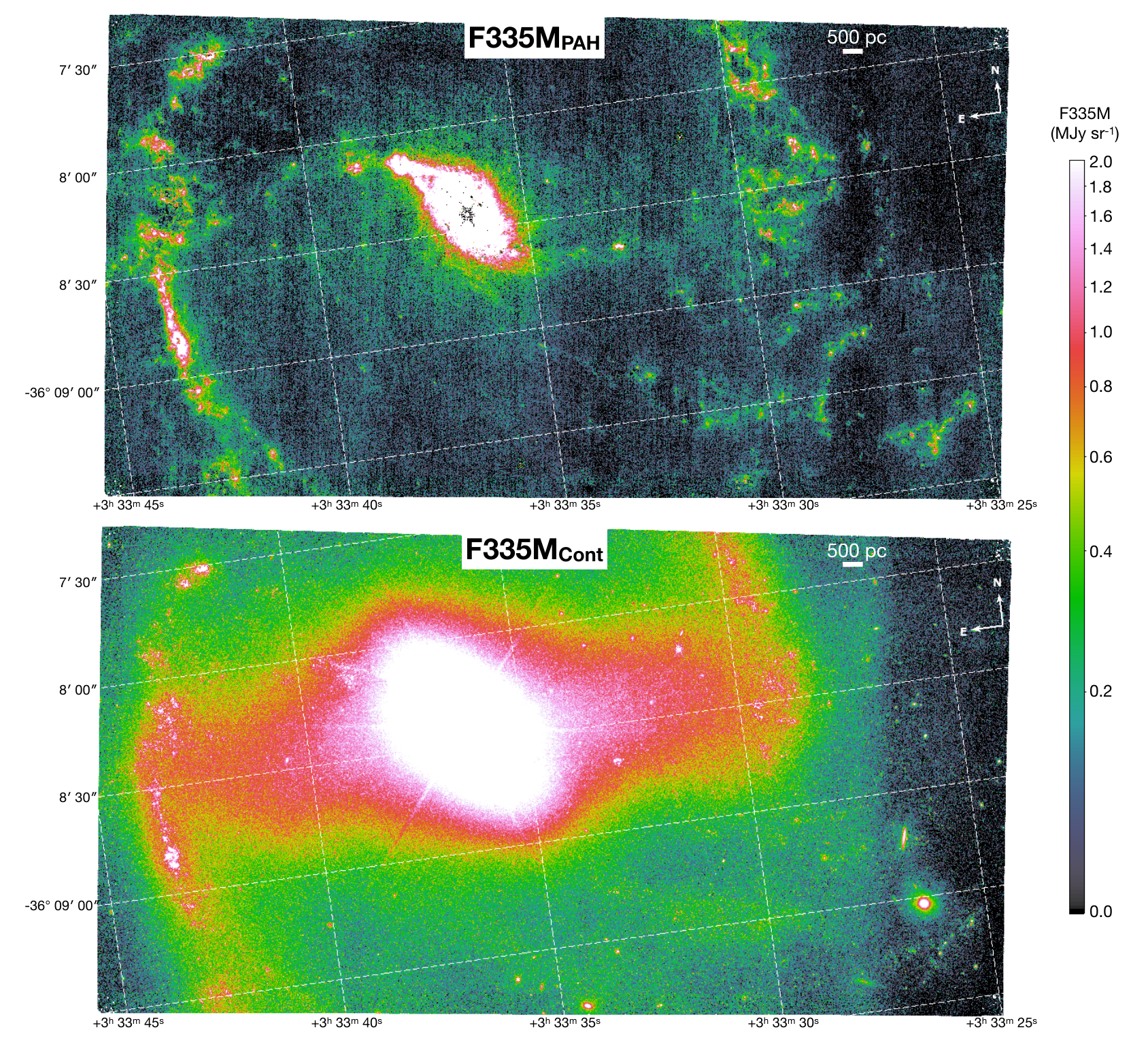}
    \caption{PAH 3.3 \micron\ emission and continuum for NGC~1365.}
    \label{fig:map_ngc1365}
\end{figure*}

\begin{figure*}
    \centering
    \includegraphics[width=\textwidth]{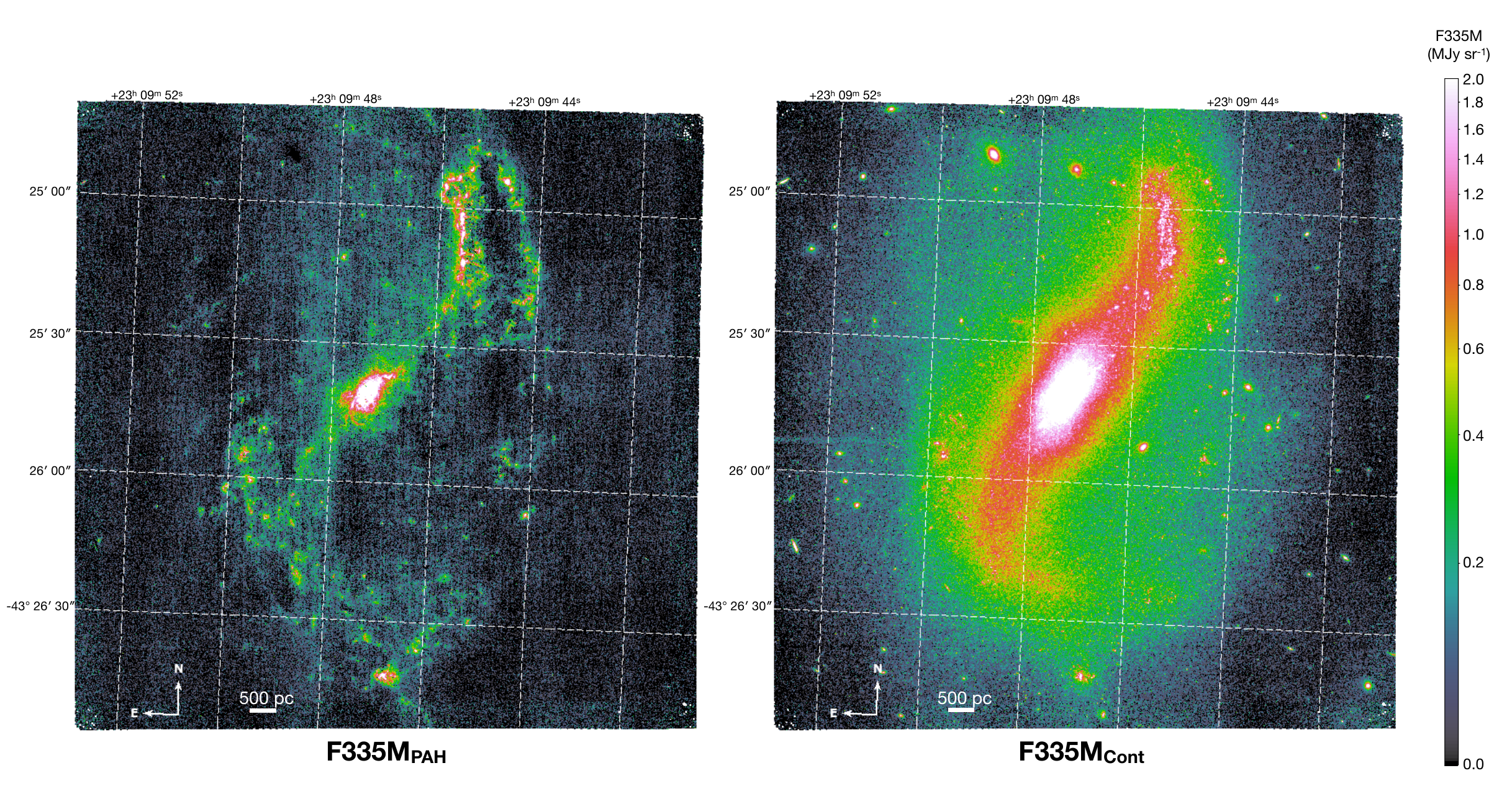}
    \caption{PAH 3.3 \micron\ emission and continuum for NGC~7496.}
    \label{fig:map_ngc7496}
\end{figure*}

In Figures~\ref{fig:map_ngc628}, \ref{fig:map_ngc1365}, and \ref{fig:map_ngc7496} we show the F335M$_{\rm PAH}$  and F335M$_{\rm cont}$ maps for all three targets, using our continuum subtraction scheme described above. The F335M$_{\rm PAH}$ maps are the highest resolution view of the PAH emission in these galaxies, with linear resolution between 5-10 pc. This resolution is similar to what can be achieved with Hubble optical imaging.  
We find typical uncertainties of $\sigma \sim 0.07$ \mjysr\ in the F335M$_{\rm PAH}$ maps for these galaxies, as measured in faint regions of the map with minimal emission (this value also matches expectations from propagating measured errors in F300M, F335M, and F360M through the correction formulae). Given that our observations required only $\sim400$ seconds of integration per field, the sensitivity of the maps is impressive. In the future, deeper observations with medium bands to map the 3.3 \micron\ PAH feature will be straightforward with NIRCam.

\subsection{PAH-to-Continuum Ratios}\label{sec:frac33}

\begin{figure*}
    \centering
    \includegraphics[width=0.45\textwidth]{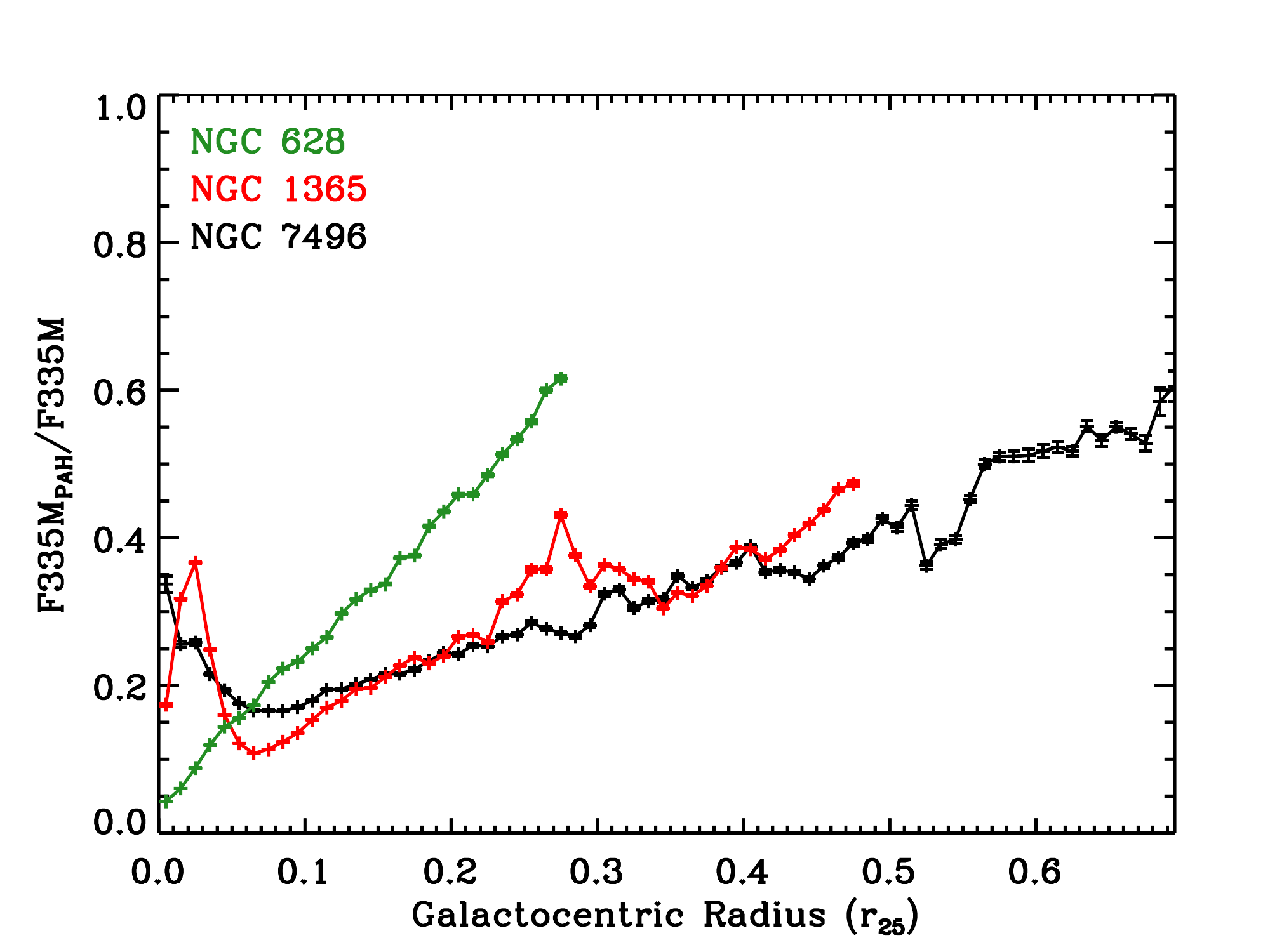}
    \includegraphics[width=0.45\textwidth]{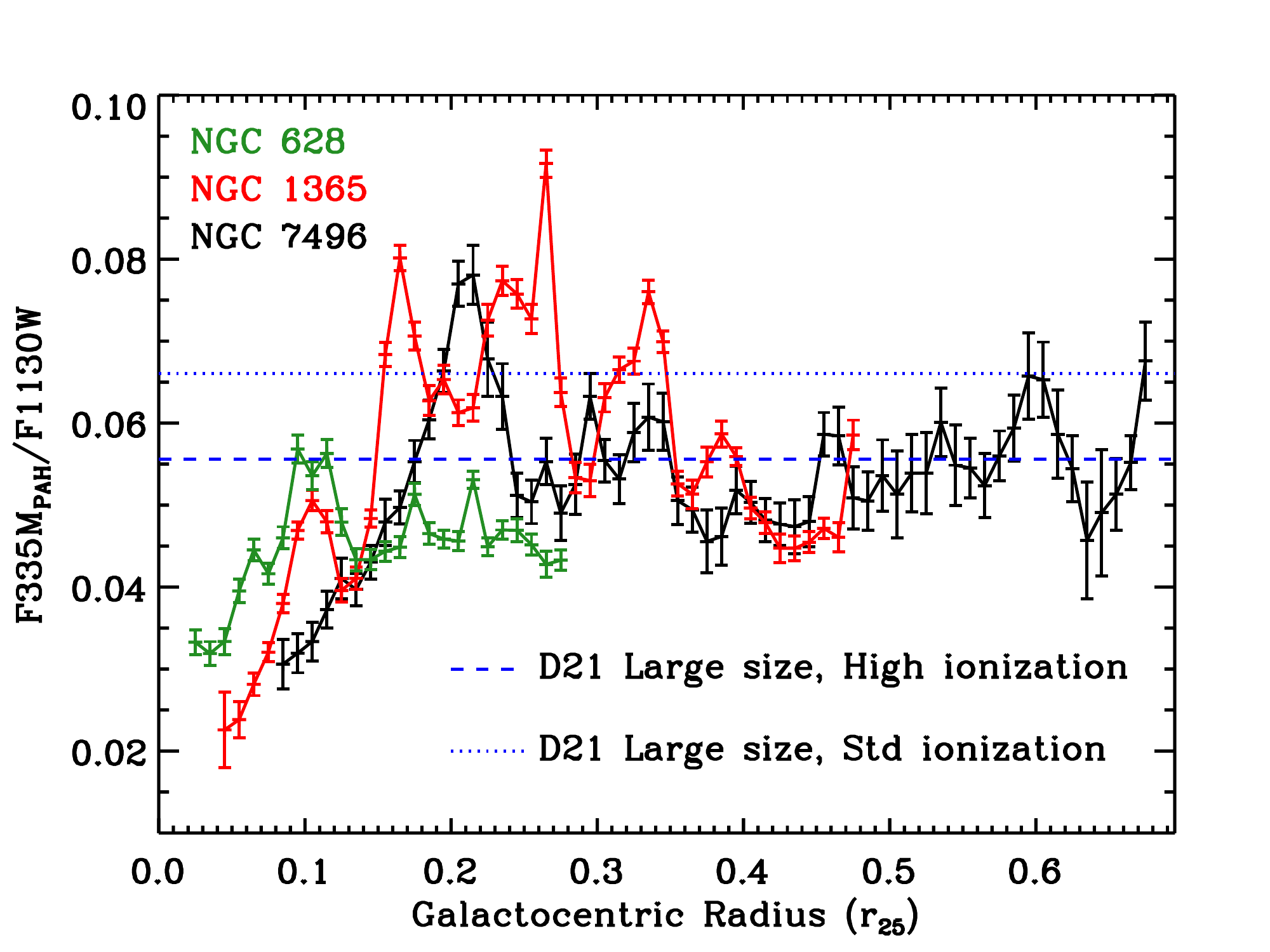}
    \caption{(left) Fraction of the F335M band from the $3.3$\micron\ PAH feature as a function of galactocentric radius in units of $r_{25}$. Error bars show the error on the mean, which is very small given the large number of measurements that contribute in each radial bin. The fraction of the F335M filter emission that traces PAHs varies systematically from the inner, stellar continuum dominated regions to the fainter outskirts of the disk. (right) The ratio of the 3.3 and 11.3 \micron\ PAH features (traced by F335M$_{\rm PAH}$/F1130W) as a function of galactocentric radius (error bars show error on the mean).  Due to saturated sources in the centers of NGC~1365 and 7496 at F1130W, we have masked the inner $r=7.5''$ region, which corresponds to the inner 0.04 r$_{25}$ for NGC 1365, and the inner 0.08 r$_{25}$ for NGC 7496.  We have also masked the inner 0.02 r$_{25}$ for NGC~628 due to contamination by evolved stellar populations at F1130W. Significant variations in the median F335M$_{\rm PAH}$/F1130W ratios as a function of radius exist in the galaxies. Our observed ratios are generally consistent with expectations from the \citet{Draine2021} models for PAH size distributions shifted towards larger grains and charge distributions with either ``high'' or ``standard'' ionization.}
    \label{fig:f335mfrac}
\end{figure*}

As a result of our continuum subtraction, we can measure the fraction of the F335M band that is due to PAH emission (i.e. F335M$_{\rm PAH}$/F335M). We show radial profiles of this fraction in Figure~\ref{fig:f335mfrac} (left). To create these profiles, we binned the F335M$_{\rm PAH}$ and F335M intensities in bins of $0.01 r_{25}$, including all pixels where the emission in both filters was detected at $>3\sigma$. We measure the median of the F335M$_{\rm PAH}$ and F335M in these bins and then divide to obtain the ratio as a function of radius. This value is lowest in the galaxy centers, where high stellar mass surface density leads to starlight dominating the F335M band. In all three galaxies the fraction increases relatively smoothly with radius outside the central 0.1$r_{25}$.  Values range between $5-65$\% , with NGC~628 spanning both the lowest and highest part of that range over the range of radii we cover ($\sim0.3$r$_{25}$). Both NGC~1365 and 7496 are barred galaxies with high central gas surface densities associated with circumnuclear star forming rings.  These regions produce  strong PAH emission, which causes the upturn in the  F335M$_{\rm PAH}$/F335M in their centers.  NGC~628, on the other hand, shows a monotonic increase in F335M$_{\rm PAH}$/F335M with radius, suggesting that starlight surface brightness falls more rapidly than PAH surface brightness as a function of radius, which may reflect varying scale-lengths for the stellar mass and ISM distributions.   The varying behavior even among these first three targets from PHANGS-JWST emphasizes the need for continuum removal recipes that work for both continuum and PAH-dominated sight-lines, since starlight makes up a highly variable fraction of the emission in the NIRCam medium bands.

\subsection{Typical 3.3/11.3 PAH Feature Ratios and Comparison to \citet{Draine2021}}

Both the 3.3 and 11.3 \micron\ PAH features are thought to arise from vibrations in C-H bonds, which are strongest from neutral grains \citep{schutte1993,vandiedenhoven2004,kerkeni2022}. Because smaller PAHs gain more energy per vibrational mode upon absorbing a UV photon of a given energy, they are able to more effectively excite the shorter wavelength emission at 3.3 \micron\ compared to larger PAHs. Thus, the 3.3/11.3 ratio for a fixed radiation field spectrum is expected to trace the average size of the grains \citep{maragkoudakis2020,Draine2021,rigopoulou2021}.  Recent models from \citet{Draine2021} predict the 3.3/11.3 ratio for a range of PAH size and charge distributions, in radiation fields of varying intensity and hardness.

In the right panel of Figure~\ref{fig:f335mfrac}, we show the radial profile of the F335M$_{\rm PAH}$/F1130W ratio for each galaxy. The data are radially binned as described in Section~\ref{sec:frac33}, with the addition of masking out radial ranges affected by saturation for the F1130W filter.  We masked the inner 7.5$''$ radius region, which extends to the radius where the point spread function for the bright central source drops to 10$^{-3}$ of its peak value \citep[see][for further details]{HASSANI_PHANGSJWST}. This corresponds to a cut at $< 0.04 r_{25}$ ($<1.4$ kpc) for NGC~1365 and $< 0.08 r_{25}$ ($<0.7$ kpc) for NGC~7496.  We additionally mask the very central region of NGC~628 ($< 0.02 r_{25}$, or $<0.3$ kpc) where evolved stellar populations contribute to the F1130W emission in a hole in the ISM distribution near the nuclear star cluster \citep[see][]{HOYER_PHANGSJWST}. Error bars show the error on the mean. Our results show variations of the ratio between $0.02-0.08$, with an average value of $\sim0.05$ across the three targets.  These ratios are investigated in more detail in \citet{CHASTENET2_PHANGSJWST} and \citet{DALE_PHANGSJWST}. For comparison, we plot two ratios from the \citet{Draine2021} models.  These values are derived by filter convolutions of the models as described in \citet{DALE_PHANGSJWST}.  We use the results for a stellar population age of 1 Gyr and representative size (``small'', ``standard'', and ``large'') and charge distributions (``low'', ``standard'', and ``high'').  For most of these combinations, the predicted F335M$_{\rm PAH}$/F1130W predictions are ouside our plot range, but our results are consistent with emission from a population of PAHs with ``large'' characteristic size and ``standard'' or ``high'' ionization.  This is in agreement with the results from \citet{DALE_PHANGSJWST} who also find the best alignment with the ``large'' and ``high'' ionization \citet{Draine2021} model grid results, although they use much younger stellar population ages to represent the environments of embedded clusters.  

\section{Conclusions} \label{sec:conc}

The capability of NIRCam medium bands on JWST to map the 3.3 \micron\ PAH feature at high angular resolution and sensitivity provides an invaluable new tool for studying PAHs. Combined with the longer wavelength MIRI imaging, the 3.3/11.3 PAH feature ratio (traced by F335M$_{\rm PAH}$/F1130W) presents one of the cleanest diagnostics of PAH size, helping to interpret a range of other band ratio variations (e.g.\ 7.7/11.3) which can have both size and charge dependence. In addition, the 3.3 \micron\ feature can be mapped with NIRCam at $2-3$ times finer angular resolution than the 7.7 \micron\ or 11.3 \micron\ bands, yielding 5--10 pc resolution in our targets. This allows measurements of the sizes of H~II regions and bubbles \citep[see][]{WATKINS_PHANGSJWST,BARNES_PHANGSJWST}, the identification of filamentary structure \citep{THILKER_PHANGSJWST,MEIDT_PHANGSJWST}, the identification of embedded clusters \citet{RODRIGUEZ_PHANGSJWST}, and potentially tracing the gas column at higher resolution than is routinely possible with any millimeter or radio facilities \citep{LEROY1_PHANGSJWST,SANDSTROM2_PHANGSJWST}.

In this Letter, we have presented a first approach to using the NIRCam medium bands F300M, F335M, and F360M to create a map of the 3.3 \micron\ PAH feature. We find a key consideration for highly resolved galaxies like our targets is to correct the F360M/F300M colors to account for contamination by PAH emission (the 3.3 \micron\ feature itself, the 3.4 \micron\ aliphatic, and 3.47 \micron\ plateau features) in F360M.  We develop an empirical first-order correction to the \citet{Lai2020} prescription, which combines the successes of the \citet{Lai2020} formula at removing starlight with a scaling using the F335M/F300M colors to correct for PAH contamination in F360M.  We demonstrate that this approach succeeds in mitigating over-subtraction of PAH emission from F335M that would result from a simple linear interpolation across the bands.  Future work to calibrate the continuum subtraction using NIRSpec observations in highly resolved nearby galaxies will be critical to move beyond our first-order correction, and deal with effects such as attenuation, absorption features, stellar population variations, hot dust, and/or nebular emission in the various bands.

%\begin{acknowledgments}
\section*{Acknowledgements}
This work is based on observations made with the NASA/ESA/CSA James Webb Space Telescope. The data were obtained from the Mikulski Archive for Space Telescopes at the Space Telescope Science Institute, which is operated by the Association of Universities for Research in Astronomy, Inc., under NASA contract NAS 5-03127 for JWST. These observations are associated with program 2107. The specific observations analyzed can be accessed via \dataset[10.17909/9bdf-jn24]{http://dx.doi.org/10.17909/9bdf-jn24}.

The authors thank the anonymous referee for feedback that improved the paper. The authors thank Thomas Lai for helpful conversations and providing fits to template spectra used in Figure~\ref{fig:spec33}. KS acknowledges funding support from grant support by JWST-GO-02107.006-A.
TGW acknowledges funding from the European Research Council (ERC) under the European Union’s Horizon 2020 research and innovation programme (grant agreement No. 694343).
JMDK gratefully acknowledges funding from the European Research Council (ERC) under the European Union's Horizon 2020 research and innovation programme via the ERC Starting Grant MUSTANG (grant agreement number 714907). COOL Research DAO is a Decentralized Autonomous Organization supporting research in astrophysics aimed at uncovering our cosmic origins.
MB acknowledges support from FONDECYT regular grant 1211000 and by the ANID BASAL project FB210003.
EJW acknowledges the funding provided by the Deutsche Forschungsgemeinschaft (DFG, German Research Foundation) -- Project-ID 138713538 -- SFB 881 (``The Milky Way System'', subproject P1). 
MC gratefully acknowledges funding from the DFG through an Emmy Noether Research Group (grant number CH2137/1-1).
FB would like to acknowledge funding from the European Research Council (ERC) under the European Union’s Horizon 2020 research and innovation programme (grant agreement No.726384/Empire)
ER and HH acknowledge the support of the Natural Sciences and Engineering Research Council of Canada (NSERC), funding reference number RGPIN-2022-03499.
KG is supported by the Australian Research Council through the Discovery Early Career Researcher Award (DECRA) Fellowship DE220100766 funded by the Australian Government. 
KG is supported by the Australian Research Council Centre of Excellence for All Sky Astrophysics in 3 Dimensions (ASTRO~3D), through project number CE170100013. 
JC acknowledges support from ERC starting grant \#851622 DustOrigin.
AKL gratefully acknowledges support by grants 1653300 and 2205628 from the National Science Foundation, by award JWST-GO-02107.009-A, and by a Humboldt Research Award from the Alexander von Humboldt Foundation.
OE gratefully acknowledges funding from the Deutsche Forschungsgemeinschaft (DFG, German Research Foundation) in the form of an Emmy Noether Research Group (grant number KR4598/2-1, PI Kreckel).

\vspace{5mm}
\facilities{JWST}

\bibliography{pah33,phangsjwst}{}
\bibliographystyle{aasjournal}

\suppressAffiliationsfalse
\allauthors

\end{document}